\documentstyle[aps,pre,twocolumn,epsf]{revtex}
\def\B.#1{{\bbox{#1}}}
\begin{document}
\title{Disentangling Scaling Properties  in Anisotropic and Inhomogeneous
Turbulence.}
 \author{Itai Arad$^{1}$, Luca Biferale$^{2,3}$, Irene Mazzitelli$^2$
   and Itamar Procaccia$^{1}$}
 \address{$^1$ Department of ~~Chemical Physics, The Weizmann Institute of
Science
Rehovot, 76100, Israel\\$^2$ Dipartimento di Fisica, Universit\`{a}
di Tor Vergata, Via della Ricerca Scientifica 1, I-00133 Roma, Italy\\
$^{3}$  INFM-Unit\'a di Tor Vergata}
\maketitle
 Version of \today
\begin{abstract}
We address scaling in inhomogeneous and anisotropic turbulent flows by
decomposing structure functions into their
irreducible representation of the SO(3) symmetry group which are designated
by $j,m$ indices. Employing simulations of channel flows with
Re$_\lambda\approx 70$ we
demonstrate that different components characterized by different $j$
display different
scaling exponents, but for a given $j$ these remain the same at different
distances
from the wall. The $j=0$ exponent agrees extremely well with high Re
measurements of the
scaling exponents, demonstrating the vitality of the SO(3) decomposition.
\end{abstract}
\vskip 0.5 cm
Most of the available data analysis and theoretical thinking about the
universal statistics of the
small scale structure of turbulence assume the existence of an idealized
model of homogeneous and
isotropic flow.  In fact most realistic flows are neither homogeneous nor
isotropic.
Accordingly, one can analyze the data pertaining to such flows in two ways.
The traditional one has been
to disregard the inhomogeneity and anisotropy, and proceed with the data
analysis assuming
that the results pertain to the homogeneous and isotropic flow. The second,
which is
advocated in this Letter, is to take the anisotropy explicitly into account,
to carefully decompose the
relevant statistical objects into their isotropic and anisotropic
contributions, and
assess the degree of universality of each component separately. We analyze
here direct numerical simulations of a channel flow with Re$_\lambda
\approx 70$
\cite{channel1,channel2,channel3}.
The main conclusion of  this Letter is that this
procedure is
unavoidable; in particular it highlights the universality
of the scaling exponents of the isotropic sector which are presumably those
governing the universal small scale statistics at very high Reynolds numbers.
In agreement with recent studies of the this subject
\cite{98ALP,98ADKLPS}we report that
different irreducible representations
of the symmetry group (characterized by indices $j,m$) exhibit scalar
functions that scale with
apparently universal exponents that differ for different $j$. The exponents
found at low values of the Reynolds number
for the $j=0$ (isotropic) sector are in excellent agreement with high Re
results; these exponents
are invariant to the position in the inhomogeneous flow, leading to
reinterpretation of recent findings of position dependence as resulting
from the
intervention of the anisotropic
sectors. The latter have nonuniversal weights that depend on the position
in the flow.

We consider here channel flow simulations on a grid of $256$ points in the
stream-wise direction
$\hat{x}$, and  $(128\times128)$ in the other two directions,
$\hat{y},\hat{z}$.
We denote by $\hat{z}$ the direction perpendicular to the walls and by
$\hat{y}$
the span-wise direction in  planes parallel to the walls. We employ periodic
boundary conditions in the span-wise and stream-wise directions and no-slip
boundary conditions on the walls. The Reynolds number based on the
Taylor scale is  $Re_{\lambda}\approx 70$ in the center of the channel
$(z=64)$.
The simulation is fully symmetric with respect to the central plane.
The flow correctly develops a mean profile in the stream-wise direction which
depends only on the distance from the wall, $U_x(z)$. The mean profile shows
the three typical regimes: a laminar linear mean profile inside the viscous
sublayers, a  logarithmic profile for intermediate distances and  finally
a parabolic mean profile  in the core of the channel.  For more details on
the averaged quantities and on the numerical code
the reader is referred to \cite{channel1,channel3}.

Previous analysis of the same data-base \cite{channel1} as well as of other
DNS \cite{steven} and experimental data \cite{exp1,exp2}
in anisotropic  flows found that the scaling properties
of energy spectra, energy co-spectra and of
longitudinal structure functions exhibit strong dependence on the local degree
of anisotropy.  For example, in  \cite{channel2}  the authors  studied  the
longitudinal structure functions at fixed distances from the walls:
$$S^{(p)}(R,z)\equiv <(v_x(x+R,y,z)-v_x(x,y,z))^p>_z$$
where $<\cdots>_z$ denotes a spatial average on a plane
at a fixed height  $z$, $1<z<64$. For this set of observables they found
that: (i)
These structure functions did not exhibit clear scaling behavior as a
function of the distance
$R$. Consequently, one needed to resort to Extended-Self-Similarity (ESS)
\cite{ess}
in order to extract a set of {\rm relative} scaling exponents
$ \hat{\zeta^z}(p) \equiv \zeta^z(p)/\zeta^z(3)$; (ii) the
relative exponents, $ \hat{\zeta^z}(p)$
depended strongly on the height $z$. Moreover,
only at the center of the channel and very close to the walls the error
bars on the relative scaling exponents extracted by using ESS were
small enough to claim the very existence of scaling behavior
in any sense.  Similarly, an experimental analysis of a turbulent
flow behind a cylinder \cite{exp1} showed a strong dependence of the
relative scaling exponents on
the  position behind the cylinder for not too big distances from
the obstacle, i.e. where anisotropic effects may still be relevant in a
wide range of scales.
In the following we present an  interpretation
of the variations in the scaling exponents observed in non-isotropic
and non-homogeneous flows upon changing the position in which the analysis
is performed. In particular, we will show that decomposing the statistical
objects into their different $(j,m)$ sectors rationalizes the findings, i.e.
scaling exponents in given $(j,m)$ sector appear quite
independent of the spatial location; only the {\em amplitudes}
of the SO(3) decomposition depend strongly on the spatial location.
These findings, if confirmed by other independent measurements, would
suggest that the apparent dependence of scaling exponents
for longitudinal structure functions on the location in a non-homogeneous
flow results from of a superposition of power laws each of which is
characterized by
its own {\em universal} scaling exponent. The amplitudes of the various
contributions
may depend on the local degree of anisotropy and non-homogeneity.

Our method of analysis is quite simple \cite{98ALP,98ADKLPS}. We start by a
direct measurement of the longitudinal structure functions
\begin{equation}
 S^{(p)}(\B.r^c,\B.R)=
\left<\left[\left(\B.u({\B.r^c + \B.R})-\B.u({\B.r^c - \B.R})\right)\cdot
\hat\B.R\right]^p \right>
\label{Spdef}
\end{equation}
Note that the two velocity fields are measured at the extremes of the diameter
of a sphere of radius $R$ centered at $\B.r^c$. Due to the inhomogeneity
this function depends explicitly on $\B.r^c$. Due to the anisotropy
the function depends on the orientation of the separation vector $2\B.R$ as
well as on its magnitude. The average must be taken over different
time frames. Typically we have used 160 time frames for
such an average. The time frames are separated by about one eddy turn over
time.
In each time frame we also improved the statistics by averaging
over one fourth of the total number of spatial points in the plane at fixed
$z$, invoking the
homogeneity in the span-wise and stream-wise directions, $\hat{x},
\hat{y}$. Thus we have
finally about  $1 \times 10^6$  contributions to each average.

Having computed $ S^{(p)}(\B.r^c,\B.R)$ we decompose it into the
irreducible representations of
the SO(3) symmetry group according to:
\begin{equation}
S^{(p)}(\B.r^c,\B.R)= \sum_{j,m} S^{(p)}_{j,m}(\B.r^c,|\B.R|) Y_{j,m}(\hat
\B.R) \ .
\label{SO3}
\end{equation}
We expect that when scaling behavior sets in (presumably at high enough Re)
we should find:
\begin{equation}
 S^{(p)}_{j,m}(\B.r^c,|\B.R|) \sim a_{j,m}(\B. r^c) |R|^{\zeta_{j}(p)} \ .
\label{scale}
\end{equation}
In other words, we expect \cite{98ALP} the scaling exponent $\zeta^j(p)$
to be independent of $m$.

The first result that we want to display is that
by applying the SO(3) decomposition we seem to improve significantly the
very existence of scaling behavior. In Fig.1 we show (i) the log-log plot
of the
raw structure function (\ref{Spdef})
with  $p=4$ measured on the central plane  with the vector ${\bf R}$
in the streamwise direction, ${\bf R} = R \hat{x}$, and (ii) the fully
isotropic sector $S^{(4)}_{0,0}(\B.r^c,|\B.R|)$ with the average in
(\ref{Spdef}) taken on the sphere centered  on the central plane $r^c_z=64$.
\begin{figure}
\epsfxsize=8truecm
\epsfysize=6 cm
\epsfbox{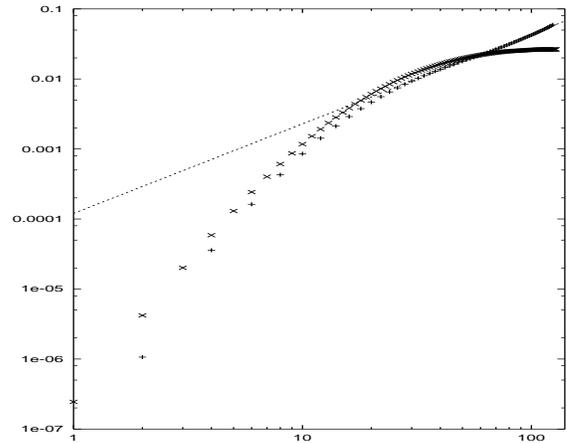}
\caption{Log-log plot of the isotropic sector of the 4th order structure
function $S^{(4)}_{0,0}$, vs. $R$ at the center of the channel $r^c_z=64$
(+). The
data represented by ($\times$) correspond to
the raw longitudinal structure function, $S^{(4)}(r^c_z=64,R\hat x)$
averaged over the central
plane only. The dashed line corresponds to the intermittent isotropic
high-Reynolds
numbers exponents $\zeta(4) = 1.28$.}
\end{figure}
It appears that already at this fairly low Reynolds number the
$j=0$ sector shows decent scaling behavior as a function of $R$. This is in
marked contrast with the raw structure function for which no scaling
behavior is
detectable ($\times$ symbols in Fig.1). For the raw quantity the method of
``extended-self-similarity" (ESS) \cite{ess} is unavoidable
if one wants to extract any kind of apparent scaling exponent. In our analysis
we found similar results also for higher order structure functions. The scaling
behavior is improved dramatically for the components and it can be seen even
without ESS. Nevertheless we will use ESS below for {\em quantitative} reading
of the exponents within every sector.

The second point we would like to stress is the apparent {\em invariance}
of the scaling
exponents belonging to the same $(j,m)$ sector with respect to changing the
spatial
location in the flow. To study this issue quantitatively we resort to ESS,
and examine the relative scaling of, say, structure functions of order
$n$ with respect to the structure function of order $2$ for $n=3,4...$.
The ESS method
is applied in each  $(j,m)$ sector separately.

In Fig.2 we show two typical ESS plots for longitudinal structure functions
of order 4 vs longitudinal structure functions of order 2 both at the center
$z=64$
and at $z=32$ in the sector $j=0$. Also, in the inset the quality of the
scaling can be appreciated by looking at the {\em logarithmic local slopes}
 of $\log(S^{(4)}_{0,0}({\bf r^c},|R|))$ vs
$\log(S^{(2)}_{0,0}({\bf r^c},|R|))$ as a function of $R$ for the same two
different central position of the sphere:
at the center of the channel ($r^c_z=64$) and at one quarter of
the total channel height ($r^c_z=32$).
The two  curves give the same
global relative scaling exponent. The best fits  for the relative scaling
exponents in the sector $(j=0)$  give
$\hat{\zeta}^{z=64}_{0}(4) \equiv
\zeta^{z=64}_{0}(4)/\zeta^{z=64}_{0}(2)=1.84 \pm
0.05$ at the center and $\hat\zeta^{z=32}_{0}(4) =1.82 \pm 0.04$
 at $r^c_z=32$. This result is remarkable and together with the experimental
result of ref.\cite{98ADKLPS} it provides strong evidence for the universality
of scaling exponent as defined in distinct  $(j,m)$ sectors. We recall
that the accepted value of this relative exponent in high-Re experiments
is $\zeta(4)/\zeta(2)\approx 1.82 \pm 0.02$ \cite{rey}.
\begin{figure}
\epsfxsize=7truecm
\epsfysize=6 cm
\epsfbox{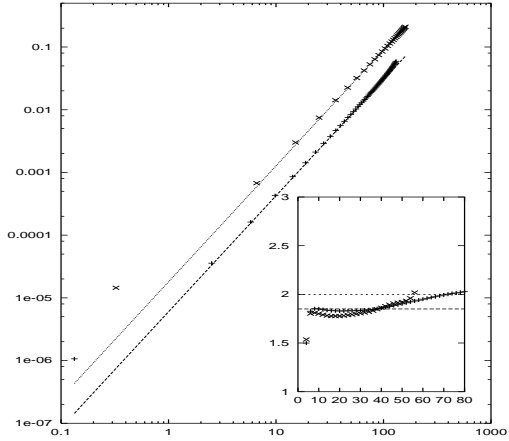}
\caption{Isotropic sector: ESS plot of $\log S^{(4)}_{0,0}$
vs $\log S^{(2)}_{0,0}$ at the center, $r^c_z=64$, (bottom curve)
 and at $r^c_z=31$ (top curve),
 the straight lines are the best fits
with slope $1.82$. Inset:  local slope  of
$\log S^{(4)}_{0,0}$, vs. $\log S^{(2)}_{0,0}$ as a function of $R$, for
two different
location in the channel: ($+$) center of the channel $r_z^c=64$, ($\times$)
one quarter of the total height $r_z^c=32$. The local slopes
are very close. For comparison we have also plotted a horizontal curve
corresponding to the accepted anomalous high-Reynolds number
value,  $\zeta(4)/\zeta(2)=1.82$.}
\end{figure}
Similarly, but affected from larger
error bars, one recovers the same invariance with respect to higher
order moments. For instance we measure $\hat{\zeta}^{z=64}_{0}(6)
\sim \hat{\zeta}^{z=32}_{0}(6) = 2.5 \pm 0.1$.  As for relative scaling
exponents of higher $j$ sectors, the scaling is less clean and therefore
we may only quote qualitative estimates. As an example,  for
relative scaling exponents of the $(j=2,m=2)$ and $(j=2,m=0)$ sectors
we have $\hat{\zeta}^{z=64}_{2,0}(4) = 1.1 \pm 0.1$,
$\hat{\zeta}^{z=32}_{2,0}(4) = 1.15 \pm 0.1$,
 $\hat{\zeta}^{z=64}_{2,2}(4) = 1.3 \pm 0.1$,
$\hat{\zeta}^{z=32}_{2,2}(4) = 1. \pm 0.1$.

To underline the quantitative improvement resulting from the
application of the SO(3) decomposition we show in Fig 3
the {\em logarithmic local slopes} of the raw structure functions
$S^{(4)}(r_z^c,R\hat{x})$ vs. $S^{(2)}(r_z^c, R \hat{x})$ at $r_z^c= 64$
and at $r_z^c= 32 $. Also the {\em logarithmic local
slopes}  of the projection on the $j=0$ sector at the same two distances
from the walls are presented. As is evident, the raw structure
function at the center of the channel and the two $j=0$ projections
are is in good agreement with the high-Reynolds numbers estimate
$\zeta(4)/\zeta(2)=1.82$ while a clearly  spurious departure is seen for the
raw structure functions at $r_c^z=32$.
\begin{figure}
\epsfxsize=8truecm
\epsfysize=6 cm
\epsfbox{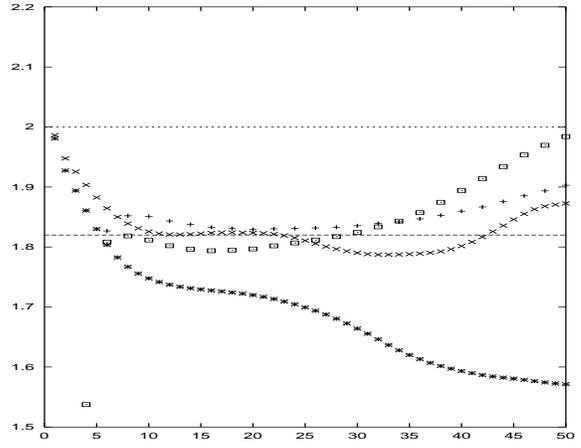}
\caption{
Logarithmic local slopes of the ESS plot of raw structure function
of order 4 versus raw structure function of order 2
at $r_c^z=64$ ($\times$), at $r_c^z=32$ ($\star$) and of the $j=0$ projection
centered at $r_c^z=64$ ($+$), and at $r_c^z=32 $($\Box$).
Also two horizontal lines
corresponding to the high-Reynolds number limit, $1.82$, and to the
K41 non-intermittent value, $2$, are shown.}
\end{figure}

Notice that due to the invariance
of Eq.(\ref{Spdef}) under the inversion ${\bf R} \rightarrow -{\bf R}$  all
the amplitudes
$a_{j,m}$ belonging to sectors with odd $j$ vanish. Similarly, at the
center of the channel
the symmetry with respect to the center $R_z \rightarrow -R_z$ forces all
the amplitudes
of the components with $j+m$ odd to vanish as well. As a consequence, the
sector $(j=2,m=1)$ is
relevant only when the center of mass is not in the central plane.
When $r^c_z=32$ we recover indeed  good scaling behavior
also for this sector but with a relative scaling exponent
$\hat{\zeta}^{z=31}_{2,1}(p)$ slightly larger
then the relative scaling exponents observed for the other $j=2$
sectors. This fact, which seems to violate the supposed foliation
in the $j$ index asserted in Eq.(\ref{scale}) is not well understood at the
moment and
it may be correlated with the presence of large scale coherent
structures (hairpin) oriented at $45^o$ with respect to the walls
observed  in all channel flow simulations \cite{moin}.

Finally, we discuss briefly the determination
of the scaling exponents associated with higher $j$ sectors. The
scaling exponent $\zeta_{j=2}(2)$ was estimated by a number of authors
on the basis of dimensional analysis \cite{lumley,yakhot,falkovich,94GLVP},
and the result is $\zeta_{j=2}(2) = 4/3$.
There is no theoretical knowledge of the actual value of this exponent
with intermittency corrections. Our direct analysis for $j=2$
seems to confirm the dimensional expectation, in agreement with the previous
experimental \cite{98ADKLPS} finding.
\begin{figure}
\epsfxsize=8truecm
\epsfysize=6 cm
\epsfbox{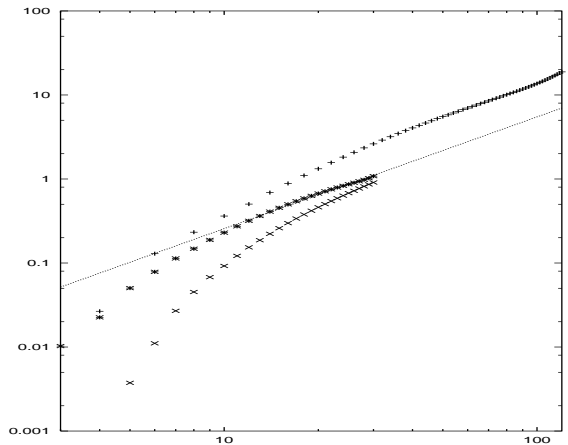}
\caption{Log-log plot of $S^{(2)}_{2,2}$ ($\star$);
$S^{(2)}_{2,0}$ ($\times$)  as
functions of $R$ at $r^c_z=32$ and of $S^{(2)}_{2,2}$ as a function of $R$
 ($+$)
 at the center, $r^c_z=64$.
The straight line correspond to the expectation of
dimensional analysis :
$\zeta_{j=2}(2)=4/3$.}
\end{figure}
In figure 4 we  show the log-log plot of  $S^{(2)}_{j,m}(\B.r^c,|R|)$
vs $|R|$  for $(j=2,m=2)$ at the center of the channel,
 and for $(j=2,m=2)$ and $(j=2,m=0)$ at $r^c_z=32$,
superimposed with the straight line with slope $4/3$.
The agreement is quite good.  Considering the relatively low Reynolds
numbers and the fact that the projections on the different sectors
depend on the non-universal prefactors $a_{j,m}$ in the decomposition
(\ref{scale}), we
think  that  together with the experimental result reported
in \cite{98ADKLPS} the present finding gives strong support
to the view that the scaling exponents in the $j=2$ sector are universal.
We are not able yet to offer the similar support to the possibility that
all the scaling exponents in the higher $j$ sectors are universal.
Such a conclusion calls for additional careful analysis of the scaling of
higher order structure
functions and higher $j$ sectors. It is outside the scope of this letter,
but it is currently under active study.

In summary, we presented three important results that follow  from the
SO(3) decomposition
of the longitudinal structure functions measured in channel flow simulations
\cite{channel1,channel3}: these are (i)
The scaling behavior is better  defined in separated $(j,m)$ sectors. This
is in contradistinction with the
raw longitudinal structure function which fails to exhibit any scaling at all.
(ii) The isotropic $(0,0)$ component of the structure functions
exhibits a universal scaling exponent which is invariant to
the spatial location in the flow and the distance from the walls.
(iii)  The $j=2$ component exhibits a scaling exponent which is compatible
with the theoretical expectation and is in excellent agreement with the
experimental
measurement \cite{98ADKLPS}, indicating universality.

The picture that emerges is that the higher order sectors are characterized
by scaling exponents that are larger than the fundamental exponent in the
isotropic sector. If this is so, it may explain the decay of anisotropy
at small scales for high Re flows. In the limit Re$\to \infty$ we expect
scaling behavior at very small values of $R/L$ with $L$ being the outer scale.
At such small scales only the smallest exponent survives, and this is
how the alleged universality of the small scales is achieved.

\section{Acknowledgments}
We are strongly indebted to F. Toschi for helping in the
set-up of the data-analysis and for a continuous and fruitful collaboration
on the subject. LB is partially supported by INFM (PRA-TURBO). IP acknowledges
the partial support of the German-Israeli Foundation, The Israel Science
Foundation, the European Commission under the Training
and Mobility of Researchers program and the  Naftali and Anna
Backenroth-Bronicki Fund for Research in Chaos and Complexity.

\end{document}